\documentclass[aps,prl,twocolumn,superscriptaddress,floatfix]{revtex4-2}
\usepackage{amssymb, amsmath,dcolumn,color,mciteplus,graphicx,subfigure}
\usepackage{calc,epsfig,epstopdf,color,mciteplus,bm,mathrsfs}
\usepackage{physics,times}
\usepackage{float}
\usepackage{lipsum}

\usepackage{xcolor}\usepackage{soul}
\definecolor{blue}{rgb}{0,0,1}
\definecolor{red}{rgb}{1,0,0}
\definecolor{green}{rgb}{0,1,0}

\usepackage[colorlinks=true, citecolor=blue, urlcolor=blue, linkcolor=black]{hyperref}
\usepackage{times}

\begin{document}

\title{Non-Abelian Aharonov-Bohm Caging 
in Synthetic Dimensions with a Trapped Ion}

\author{Wanchao Yao}\email{These authors contribute equally.}
\affiliation{Laboratory of Spin Magnetic Resonance, School of Physical Sciences, Anhui Province Key Laboratory of Scientific Instrument Development and Application, University of Science and Technology of China, Hefei, 230026, China}

\author{Sai Li}\email{These authors contribute equally.}
\affiliation{Key Laboratory of Atomic and Subatomic Structure and Quantum Control (Ministry of Education),\\ Guangdong Basic Research Center of Excellence for Structure and Fundamental Interactions of Matter,\\ and School of Physics, South China Normal University, Guangzhou 510006, China}
\affiliation{Guangdong Provincial Key Laboratory of Quantum Engineering and Quantum Materials, Guangdong-Hong Kong Joint Laboratory of Quantum Matter, and Frontier Research Institute for Physics,\\ South China Normal University, Guangzhou 510006, China}

\author{Zhiyuan Liu}\email{These authors contribute equally.}
\affiliation{Laboratory of Spin Magnetic Resonance, School of Physical Sciences, Anhui Province Key Laboratory of Scientific Instrument Development and Application, University of Science and Technology of China, Hefei, 230026, China}

\author{Yi Li}
\affiliation{Laboratory of Spin Magnetic Resonance, School of Physical Sciences, Anhui Province Key Laboratory of Scientific Instrument Development and Application, University of Science and Technology of China, Hefei, 230026, China}
\affiliation{National Advanced Talent Cultivation Center for Physics, University of Science and Technology of China, Hefei, 230026, China}
\affiliation{Hefei National Laboratory, Hefei 230088, China}

\author{Zihan Xie}
\affiliation{Laboratory of Spin Magnetic Resonance, School of Physical Sciences, Anhui Province Key Laboratory of Scientific Instrument Development and Application, University of Science and Technology of China, Hefei, 230026, China}
\affiliation{Hefei National Laboratory, Hefei 230088, China}

\author{Xingyu Zhao}
\affiliation{Laboratory of Spin Magnetic Resonance, School of Physical Sciences, Anhui Province Key Laboratory of Scientific Instrument Development and Application, University of Science and Technology of China, Hefei, 230026, China}
\affiliation{Hefei National Laboratory, Hefei 230088, China}

\author{Xu Cheng}
\affiliation{Laboratory of Spin Magnetic Resonance, School of Physical Sciences, Anhui Province Key Laboratory of Scientific Instrument Development and Application, University of Science and Technology of China, Hefei, 230026, China}
\affiliation{Hefei National Laboratory, Hefei 230088, China}

\author{Yue Li}
\affiliation{Laboratory of Spin Magnetic Resonance, School of Physical Sciences, Anhui Province Key Laboratory of Scientific Instrument Development and Application, University of Science and Technology of China, Hefei, 230026, China}

\author{Zheng-Yuan Xue}\email{zyxue83@163.com}
\affiliation{Key Laboratory of Atomic and Subatomic Structure and Quantum Control (Ministry of Education),\\ Guangdong Basic Research Center of Excellence for Structure and Fundamental Interactions of Matter,\\ and School of Physics, South China Normal University, Guangzhou 510006, China}
\affiliation{Guangdong Provincial Key Laboratory of Quantum Engineering and Quantum Materials, Guangdong-Hong Kong Joint Laboratory of Quantum Matter, and Frontier Research Institute for Physics,\\ South China Normal University, Guangzhou 510006, China}
\affiliation{Hefei National Laboratory, Hefei 230088, China}

\author{Yiheng Lin} \email{yiheng@ustc.edu.cn}
\affiliation{Laboratory of Spin Magnetic Resonance, School of Physical Sciences, Anhui Province Key Laboratory of Scientific Instrument Development and Application, University of Science and Technology of China, Hefei, 230026, China}
\affiliation{Hefei National Research Center for Physical Sciences at the Microscale, University of Science and Technology of China, Hefei 230026, China}
\affiliation{Hefei National Laboratory, Hefei 230088, China}

\date{\today}

\begin{abstract}

Aharonov-Bohm (AB) caging is a complete localization phenomenon in two-dimensional lattices due to destructive interference induced by the background gauge fields. However, current investigations of AB caging are mostly restricted to the Abelian gauge field case, and the observation of AB caging under non-Abelian gauge fields in a quantum system still remains elusive.
Here, we report experimental realization of tunable synthetic non-Abelian $SU(2)$ gauge fields in a rhombic lattice, engineered within the synthetic dimensions of a vibrating trapped ion with multiple levels. We realize AB caging under both Abelian and non-Abelian gauge fields and systematically investigate the distinctive transport properties of the non-Abelian case. In particular, we observe typical emergent quantum dynamics unique to non-Abelian AB caging, including initial-state-dependent dynamics, second-order effects, and asymmetric caging behavior. These observations demonstrate the trapped ion system as a powerful platform for simulating  emergent phenomena in high-dimensional quantum systems with exotic synthetic gauge fields.

\end{abstract}

\maketitle

\textit{Introduction}---The gauge field theory is the cornerstone of modern physics \cite{Paulson21, Hali25, YLiu25, Meth25}.
Since being introduced by Yang and Mills \cite{YM54}, the non-Abelian gauge theory has played an important role in illustrating exotic quantum dynamics \cite{Sundrum86, Hauke12, NAChen19, NAYou22, NAYang24}.
In recent years, due to the fast development of quantum technologies, it is possible to artificially synthesize the gauge field to explore quantum many-body physics with emergent states of quantum matter \cite{Lin09, Aidel11, Aidel13, Tzuang14, An17, Zhou23, QLin23, YLi23, Rosen24, Wang24, YLiuAppl25}. The Aharonov-Bohm (AB) effect \cite{AB59, Fang12, Noguchi14, Kawaguchi24} is one of the most profound examples, which predicts the cumulative geometric phase of the wave packet of a charged particle moving under the background gauge field. Under elaborate lattice geometry and specially chosen magnetic flux, the AB effect leads to destructive interference among different transport paths, and thus results in a complete localization, that is, the AB caging phenomenon \cite{Vidal98,Abilio99}.

As the paradigm for the study of transport and localization \cite{Bermudez11, Liberto19, Gli19, Gli20, Aravena22, Zhang25, Chen25}, AB caging has been demonstrated on various platforms, including trapped ions \cite{Saner25}, Rydberg atoms \cite{Chen24, Chen25}, photonic lattice \cite{Muk18, Kremer20, Wang25}, and ultracold atoms \cite{Li22}. However, these works have been mostly restricted to the case with Abelian gauge field. Recently, the theory of non-Abelian AB caging has been proposed \cite{Li20, CPL25} and experimentally explored in a classical system \cite{Zhang23}. Under the non-Abelian gauge field \cite{Neef23, Wong25, Pang24, Dong24, Osterloh05, Goldman14, Liang24, Yang19, Cheng25}, the non-Abelian AB caging is shown to host significantly distinct dynamical properties compared to its Abelian counterpart, such as second-order effects and the asymmetric caging phenomenon. However, the observation of non-Abelian AB caging in a quantum system still remains elusive, due to the requirement of a multi-level simulator and high precision coherent manipulation. As one of the leading platforms for quantum simulation, trapped ions naturally possess a multi-level structure \cite{Ringbauer22, Hrmo23, Meth25}, making it an ideal platform for the simulation of complex lattice structures. In addition, the quantized levels of the trapped ion motion, named below interchangeably as phonons, can be regarded as the synthetic dimensions \cite{Wang24}, which alleviates experimental challenges in spatial lattices. Therefore, it provides a programmable and scalable way to explore high-dimensional quantum physics \cite{Celi14, Lustig19, Yuan21, Liang24, Chen24, Luengo24}.

In this Letter, we experimentally synthesize the $SU(2)$ gauge field in an effective rhombic lattice. The lattice is constructed in the spin-phonon synthetic dimensions of a single ${}^{40}{\rm Ca}^+$ ion confined in a Paul trap, and the target gauge field is achieved by tuning the magnitude and phase of the driving laser fields.
{We first synthesize the Abelian and the non-Abelian gauge fields, and realize the AB caging for both cases. Then,  we demonstrate the unique dynamical properties of the non-Abelian AB caging, including initial-state dependence, the second-order caging and the asymmetric caging phenomena. Furthermore, by tuning the phases of coupling lasers, we observe the $SU(2)$ interference dynamics with the preservation of population under the non-Abelian gauge field.  These results lay the foundation for future exploration of high-dimensional quantum physics and emergent new states of quantum matter in presence of non-Abelian gauge fields.}

\begin{figure}[t]
\centering
\includegraphics[width=0.95\linewidth]{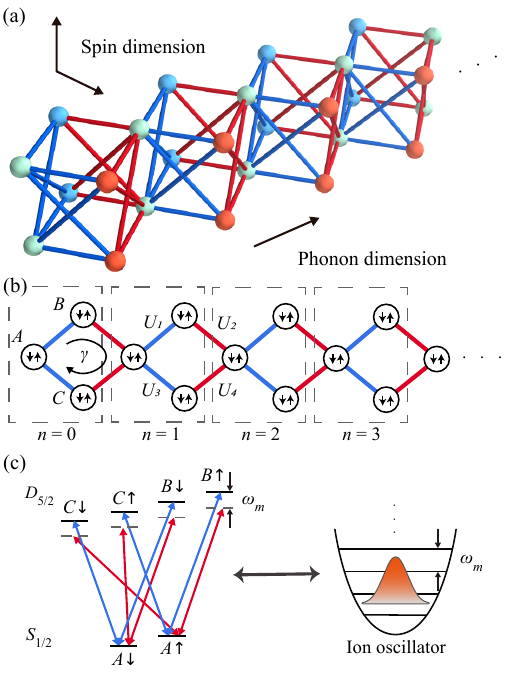}
\caption{Synthesis of $SU(2)$ gauge fields in the spin-phonon synthetic dimensions. (a) The two-layer rhombic lattice. Each lattice site is encoded in a spin level with a specific phonon number $n$. The inter-site couplings are realized via couplings between the internal spin states (blue bonds) and the spin-motion couplings (red bonds). (b) The effective rhombic lattice with internal spins. The gauge field $U_i\in SU(2)$ is synthesized on the link by controlling the magnitude and the phase of the couplings. The Wilson loop is defined by $|{\rm Tr}[U(\gamma)]|:=|{\rm Tr}(U_3U_4U_2U_1)|$ of loop $\gamma$ in a plaquette. (c) Encoding the lattice into a ${}^{40}{\rm Ca}^+$ ion. Four levels in the $D_{5/2}$ manifold are used to encode sites B (blue balls in (a)) and sites C (orange balls). Two levels in the $S_{1/2}$ manifold are used to encode sites A (light green balls). Four lasers resonant with the quadrupole transition drive the carrier transition (blue arrows). Another four lasers detuned by $-\omega_m$ drive the red sideband transition (red arrows).}
\label{fig1}
\end{figure}

\textit{Model and experimental setup}---To synthesize the $SU(2)$ gauge fields, we consider a two-fold rhombic lattice system, as shown in Fig.~\ref{fig1}(a), synthesized from the spin-phonon dimensions of the vibrating ion. Each lattice site is encoded in a spin-$3/2$ level $\Xi_j$ with
internal electronic levels, where $\Xi\in\{A,B,C\}$ and $j\in\{\uparrow,\downarrow\}$, spanned by the phonon number $n$ of the motional mode. The lattice sites are connected with mutual or neighboring phonon numbers via laser couplings
with tunable phase and amplitude.
The $SU(2)$ gauge fields are thus synthesized in the effective lattice as shown in Fig.~\ref{fig1}(b), which is a projection of the original lattice.
The system can be described by an effective Hamiltonian
\begin{eqnarray}
H&=&\frac{J}{2}\sum_{n\geq 0}b_n^\dagger U_1a_n+c_n^\dagger U_3a_n+a^\dagger_{n+1}U_2b_n\nonumber\\
&&+a^\dagger_{n+1}U_4 c_n+{\rm h.c.},
\end{eqnarray}
where $U_i\in SU(2)(i=1,2,3, {\rm and\ }4)$ is the synthesized inter-site gauge field, $J$ is the coupling strength and $\xi_n$ ($\xi \in \{a,b,c\}$) is the particle annihilation operator of site $\Xi_n$, with $\xi_n=(\xi_{n,\downarrow}, \xi_{n,\uparrow})^T$.

The $SU(2)$ gauge fields manifest unique non-Abelian AB caging effects, characterized by the Wilson loop $|{\rm Tr}[U(\gamma)]|:=|{\rm Tr}(U_3U_4U_2U_1)|<2$ \cite{Goldman09, Goldman14}, see Fig.~\ref{fig1}(b), where the transport is spin-dependent. We consider a particle initialized at site $A_n$ with spin state $|\psi\rangle$, which experiences interference dynamics under $H$. Hopping from $A_n$ to $A_{n+1}$ is characterized by the interference matrix
\begin{equation}
T=\frac{1}{2}(U_2U_1+U_4U_3),
\end{equation}
owing to the interference between unitary transformation $U_2U_1$ accumulated along the upper path and $U_4U_3$ along the lower path.
Similarly, the hopping from $A_n$ to $A_{n-1}$ is characterized by the Hermitian conjugate $T^\dagger$. A special case is the destructive interference under the condition $T^m|\psi\rangle=T^{\dagger n}|\psi\rangle=0$, which leads to the $m^{\rm th}$-order rightward ($n^{\rm th}$-order leftward) AB caging phenomenon.
Here, we focus on demonstration of following aspects: (1    ) the initial-state $|\psi\rangle$ dependence for the rightward caging condition $T^m|\psi\rangle=0$, (2) the rightward second-order caging with $m=2$,
(3) the asymmetry between the rightward and leftward transport due to $T\neq T^\dagger$.

\begin{figure*}[t]
\centering
\includegraphics[width=0.95\linewidth]{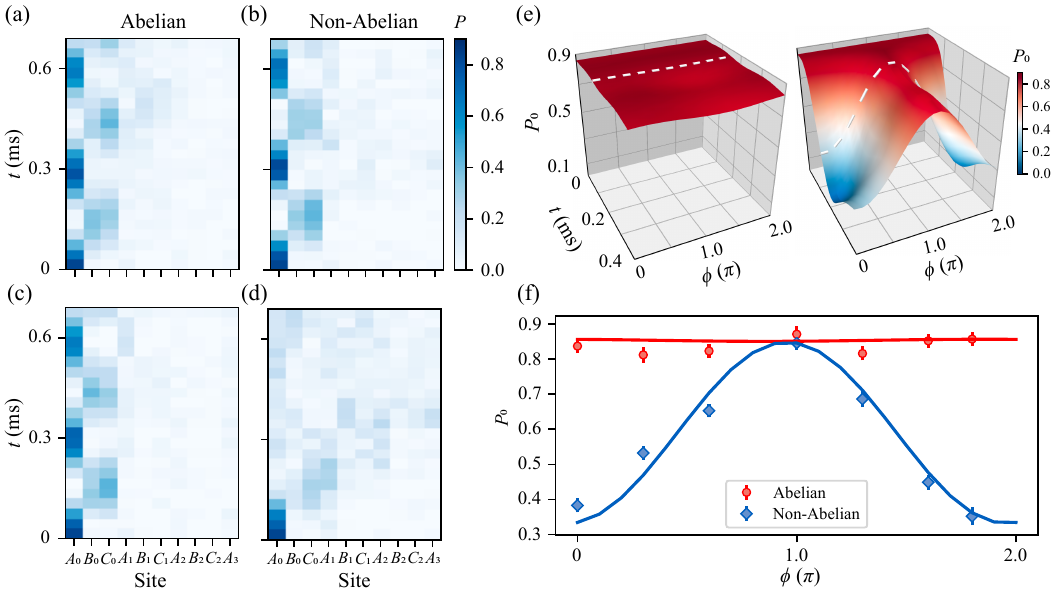}
\caption{Realization of AB caging. (a-d) Dynamical evolution diagrams under the Abelian (left panel) and the non-Abelian gauge field (right panel). We initialize the system with an out-of-phase superposition state (a)(b) and an in-phase superposition state (c)(d) at site $A_0$, respectively. (e)(f) The initial-state dependent dynamics. We continuously tune the relative phase $\phi$ of the initial state. $P_0$ denotes the sum of the population of all the $n=0$ sites. (e) Numerical simulated plot of $P_0$ versus time $t$ and the initial relative phase $\phi$ for the Abelian (left panel) and the non-Abelian gauge field (right panel). (f) The experiment is carried out at the time of $t=0.15$ ms (dashed lines in (e)). The orange dots (blue diamonds) are the experimental results for the Abelian (non-Abelian) case. The theory lines are as in (e). Error bars stand for 1 standard deviation.}
\label{fig2}
\end{figure*}

Experimentally, we implement the lattice in a single ${}^{40}{\rm Ca}^+$ ion confined in a Paul trap \cite{bible98}, utilizing the coupled multiple internal levels and external quantized motion (Fig.~\ref{fig1}(c)). The spin-$3/2$ level is encoded in four levels with $m_j=\{\pm3/2,\pm1/2\}$ in the $D_{5/2}$ manifold and two levels in the $S_{1/2}$ manifold \cite{NC25}. The secular motion with frequency $\omega_m\approx2\pi\times1.09\ {\rm MHz}$ is utilized as the phonon dimension in our experiment.
We use four resonant narrow-linewidth 729 nm lasers to drive the couplings between the internal spin states, and another four lasers detuned by $-\omega_m$ to drive the spin-motion couplings, which connect lattice sites with the neighboring phonon number. We tune the amplitude and the phase of the lasers to realize the desired gauge field $U_i$, with the coupling strength $J=2\pi\times2.5{\rm\ kHz}$ throughout our experiment. Note that our model is not perfectly translational-invariant due to the extra factor $\sqrt{n+1}$ for the red sideband transition \cite{Leibfried03}. However, we expect no influence for the interference phenomena for such an extra factor, since it simultaneously acts on both the upper and lower path of a plaquette, and thus becomes  a global factor for the interference matrix.

At the beginning of the experiment, we first cool the ion motion to near the ground state and prepare the ion to the target spin-motion state by laser pulses as detailed in following sections. At the end of the evolution, a joint spin-motion detection is performed for readout \cite{sm}.

\textit{Initial-state dependence of AB caging}---We first demonstrate the AB caging under a special initial state, such that the Abelian and the non-Abelian gauge field behave similarly. For the Abelian case, we set 
$$U_1=U_4=\begin{bmatrix}0 & 1\\1 & 0\end{bmatrix}, \quad U_2=U_3=\begin{bmatrix}-1 & 0\\0 & 1\end{bmatrix},$$ 
which yields an interference matrix with $T=\begin{bmatrix}0 & 0\\0 & 0\end{bmatrix}$, corresponding to Wilson loop $|{\rm Tr}[U(\gamma)]|=2$.
We initialize the system at $A_0$ with out-of-phase superposed state $|\psi_{\rm o}\rangle=\frac{1}{\sqrt{2}}(-1, 1)^T$. After the  evolution with duration $t$, the measured population is shown in Fig.~\ref{fig2}(a). We observe a dynamics mostly restricted between $A_0,B_0,C_0$ attributed to the destructive interference, with minor leakage to the $n>0$ sites from the experimental imperfection, thus demonstrating the phenomenon of Abelian AB caging. Next, we turn to the non-Abelian case of
$$U_1=U_2=U_3=\begin{bmatrix}1 & 0\\0 & 1\end{bmatrix}, \quad U_4=\begin{bmatrix}0 & 1\\1 & 0\end{bmatrix}.$$ 
The interference matrix $T=\frac{1}{2}\begin{bmatrix}1 & 1\\1 & 1\end{bmatrix}$ corresponds to a zero Wilson loop. With the same initial state as in the Abelian case, we observe the dynamics, as shown in Fig.~\ref{fig2}(b), where the population is mostly restricted to the $n=0$ sites as expected.

\begin{figure*}[t]
\centering
\includegraphics[width=\linewidth]{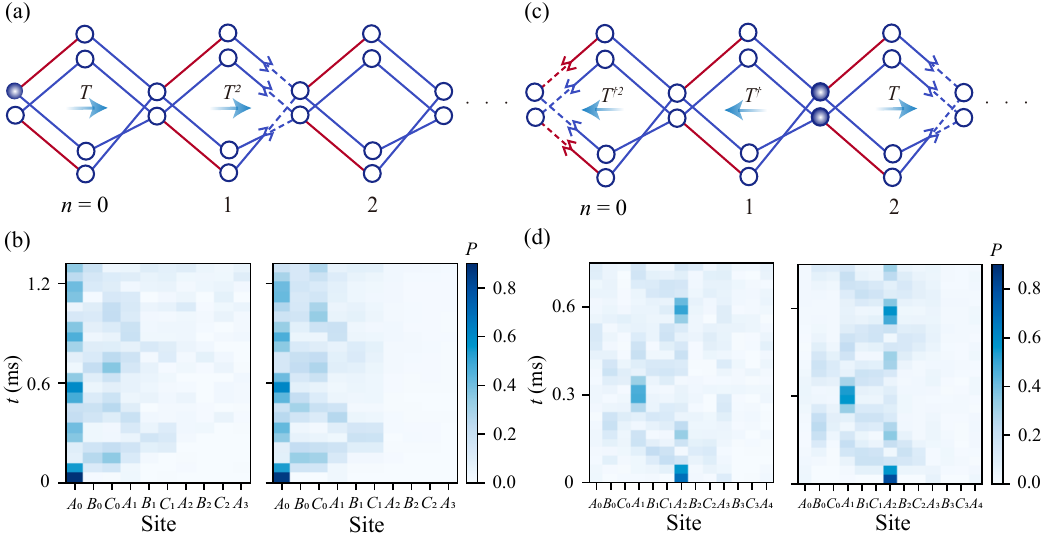}
\caption{Unique dynamics of non-Abelian AB caging. (a) Illustration of the second-order caging. The solid ball indicates the initial state. The red (blue) lines are the couplings with phase $\pi$ (0). (b) The experimental results (left) and the numerical simulations (right) show that the population localizes inside the $n\leq1$ sites. (c) Illustration of the asymmetry caging. The colors of the lines are set in the same way as in (a). (d) The experimental results (left) and the numerical simulations (right) show that the destructive interference occurs at $A_0$ and $A_3$.}
\label{fig3}
\end{figure*}

To investigate the initial-state dependence of the non-Abelian gauge field case, with the same $U_i$ as above, we instead set the initial state to be in-phase superposed state 
$|\psi_{\rm in}\rangle=\frac{1}{\sqrt{2}}(1, 1)^T$
at $A_0$.
As shown in Fig.~\ref{fig2}(c)(d) for the experimental implementation, while the caging phenomenon is still observed for the Abelian coupling, the population quickly spreads out for the non-Abelian coupling, distinct from the initial state of $|\psi_{\rm o}\rangle$ shown above. To illustrate such a transition, in Fig.~2(e), we numerically simulate the dynamics for various $\phi$ starting from a generic state $|\psi\rangle=\frac{1}{\sqrt{2}}(e^{i\phi}, 1)^T$ including the experimental imperfection \cite{sm}, and sum the population for all the $n=0$ sites, denoted by $P_0$, showing the stark contrast between the two cases.
We experimentally verify such predictions by measuring $P_0$ at the evolution duration of $t=0.15$ ms, as shown in Fig.~\ref{fig2}(f). For the Abelian coupling, we observe high population for all $\phi$ we choose. For the non-Abelian coupling, the measured $P_0$ reaches the maximum at $\phi=\pi$ and decreases significantly as $\phi$ changes away from $\pi$, in agreement with numerical results.
Thus, we verify the initial-state dependence for the caging phenomenon. In the following, we will focus on the non-Abelian AB caging and describe its unique dynamics.

\textit{Dynamics of non-Abelian AB caging}---To demonstrate the second-order caging phenomenon with $T^2=0$, we set 
$$U_1=\begin{bmatrix}-1 & 0\\0 & 1\end{bmatrix},U_2=\begin{bmatrix}1 & 0\\0 & 1\end{bmatrix},U_3=\begin{bmatrix}1 & 0\\0 & -1\end{bmatrix},U_4=\begin{bmatrix}0 & 1\\1 & 0\end{bmatrix},$$ 
which corresponds to $T=\frac{1}{2}\begin{bmatrix}-1 & -1\\1 & 1\end{bmatrix}\neq0$. This dynamics is illustrated in Fig.~\ref{fig3}(a), where the state starting from $|\psi\rangle=(1, 0)^T$ at $A_0$ would propagate towards higher $n$, but destructively interfere at $A_2$; 
in addition, we observe a negligible population in the $\xi_{n\geq2}$ sites with $\xi=A,B,C$. These measurements demonstrating the second-order caging phenomenon, agree with numerical simulations in Fig.~\ref{fig3}(b).

To realize asymmetry caging,
we set the same coupling as in the second-order caging experiment, with $$U_1=\begin{bmatrix}-1 & 0\\0 & 1\end{bmatrix},U_2=\begin{bmatrix}1 & 0\\0 & 1\end{bmatrix},U_3=\begin{bmatrix}1 & 0\\0 & -1\end{bmatrix},U_4=\begin{bmatrix}0 & 1\\1 & 0\end{bmatrix},$$ but instead initialize the system as $|\psi_{\rm o}\rangle=\frac{1}{\sqrt{2}}(1,  -1)^T$ at $A_2$. The dynamics is illustrated in Fig.~\ref{fig3}(c). Similarly to the previous experiment, the second-order caging phenomenon is expected for the leftward transport. However, the initial state we choose changes the behavior of the rightward transport since $T|\psi\rangle=0$ is satisfied. Thus, we expect the first-order caging to proceed rightwards. We carry out the experiment along with the numerical simulations as shown in Fig.~\ref{fig3}(d).
After evolution, we observe that the wave packet moving rightward and that moving leftward destructively interfere at $A_3,A_0$, respectively. Then the reflected wave packets constructively interfere at $A_1$, resulting in a strong signal at about $t=0.3$ ms. A similar strong interference signal arises again at $A_2$ at a later time. These strong signals, together with the low population at $A_0$ and $A_3$ during evolution, form evidence of asymmetric caging, thus verifying our assertion.

\begin{figure}
\centering
\includegraphics[width=0.95\linewidth]{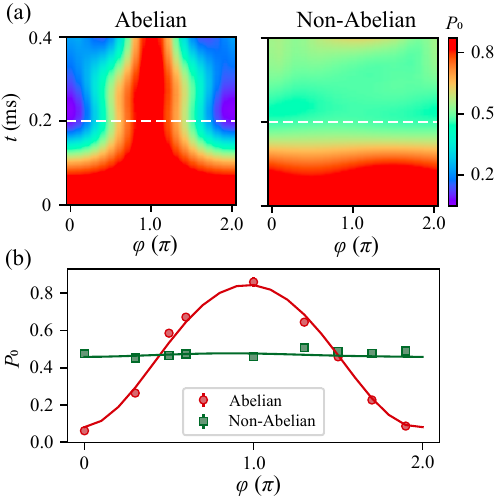}
\caption{$SU(2)$ interference dynamics. We simultaneously tune the phase $\varphi$ of two couplings. (a) Numerical calculated dynamics of $P_0$. (b) Experimental results at the time of $t=0.2\ {\rm ms}$ (dashed lines in (a)). The red circles (green squares) are the results of the Abelian (non-Abelian) gauge fields. The theory lines are as in (a). Error bars stand for 1 standard deviation, and are smaller than markers.}
\label{fig4}
\end{figure}

\textit{$SU(2)$ interference dynamics}---Finally, to fully investigate the $SU(2)$ interference dynamics, we study the dynamics under different coupling phases.
The couplings are set as $$U_1=U_4=\begin{bmatrix}0 & 1\\1 & 0\end{bmatrix}, \quad U_2=U_3=\begin{bmatrix}e^{i\varphi} & 0\\0 & 1\end{bmatrix}$$
to realize the Abelian field and $$U_1=\begin{bmatrix}e^{i\varphi} & 0\\0 & 1\end{bmatrix},U_2=\begin{bmatrix}1 & 0\\0 & 1\end{bmatrix},U_3=\begin{bmatrix}1 & 0\\0 & e^{i\varphi}\end{bmatrix},U_4=\begin{bmatrix}0 & 1\\1 & 0\end{bmatrix}$$
to realize the non-Abelian field with tunable $\varphi$.
We initialize the system with $|\psi\rangle=(1, 0)^T$ at $A_0$ and sum the population of the $n=0$ sites to obtain $P_0$ at the end of the evolution. Interestingly, the numerical simulations in Fig.~\ref{fig4}(a) show that both Abelian and non-Abelian coupling behave significantly different at $\varphi=0$, which corresponds to perfectly constructive interference and free quantum walk for both cases. We experimentally verify the predictions at the time
of $t=0.2\ {\rm ms}$, as shown in Fig.~\ref{fig4}(b). We observe that with the same evolution time, the population of the Abelian coupling drops to near zero, while that of the non-Abelian coupling remains relatively high. We attribute the preservation to its complex coupling, where the all-to-all connection enables the wave packet to interfere and circulate inside the $n=0$ lattice sites, before it finally spreads out. The measured dynamics for other $\varphi$ we choose is also in agreement with numerical simulations.
Note that the non-Abelian field exhibits second-order caging under this coupling, and we avoid the sum of the population of too many sites considering the experimental precision. Thus a high population around $\varphi=\pi$ is absent.

\textit{Conclusion}---In summary, we synthesized the $SU(2)$ gauge fields in the rhombic lattice constructed in the spin-phonon synthetic dimensions of a single trapped ion. By tuning the couplings, we synthesize both Abelian and non-Abelian gauge fields. We realize AB caging under both cases, and observe the initial-state dependence only for the latter. We investigate the exotic dynamical properties of the non-Abelian AB caging through the realization of the second-order caging and the asymmetry caging. We further explore the dynamics under different coupling phases and observe the population preservation for the non-Abelian gauge field. Our results pave the way for future exploration of high dimensional gauge fields \cite{Dong24}. The synthesized rhombic lattice could potentially be applied for detecting magnetic flux in quantum precision measurement \cite{Brosco21}.
Besides, our methods can be generalized to multiple ions, where the interaction-driven dynamics may shed light on the study of transport \cite{Chen25}, localization \cite{Nicolau23, Santos14} and topological insulators \cite{Wei24, Kremer20}.

This work was supported by the National Natural Science Foundation of China (Grants No. 92565306, No. 92576110, No. 12275090, No. 12304554 and No. 92165206), the Chinese Academy of Sciences (Grant No.~XDB1300000) and Quantum Science, National Key Research and Development Program of China (Grant No. 2025YFE0217900), and Technology-National Science and Technology Major Project (Grants No. 2021ZD0301603 and No. 2021ZD0302303).

\clearpage

\begin{center}
\textbf{Supplementary Material: Non-Abelian Aharonov-Bohm Caging in Synthetic Dimensions with a Trapped Ion}
\end{center}

\setcounter{figure}{0}
\setcounter{equation}{0}
\renewcommand*{\thefigure}{S\arabic{figure}}
\renewcommand*{\theequation}{S\arabic{equation}}

The Supplementary Information provides details about (I) the experimental Hamiltonian, (II) experimental setup, data processing, (III) measurement of the Wilson loop, (IV) numerical simulation, and (V) analysis of the caging condition.

\section{The experimental Hamiltonian}

The Hamiltonian we want to simulate has the following form
\begin{eqnarray}
H&=&\frac{J}{2}\sum_{n\geq0}(b_n^\dagger U_1a_n+c_n^\dagger U_3a_n\nonumber\\
&&+a^\dagger_{n+1}U_2b_n
+a^\dagger_{n+1}U_4 c_n)+{\rm h.c.},
\label{eq1}
\end{eqnarray}
which describes the particles hopping in a rhombic lattice under the background gauge field $U_i\in SU(2)$. It can be realized by the Hamiltonian of an ion trapped in the harmonic potential with the ion-laser interaction. The energy of a two-level ion trapped in harmonic confinement is described by the Hamiltonian
\begin{equation}
H_0=\frac{\omega_0}{2}\sigma_z+\omega_m(a^\dagger a+\frac{1}{2}),
\end{equation}
where $a$ is the annihilation operator of the quantum oscillator, $\omega_0$ is the unperturbed two-level transition frequency and $\omega_m$ is the motional frequency. A laser with frequency $\omega_L$ and phase $\phi_L$ induces the interaction described by the Hamiltonian 
\begin{eqnarray}
H_{L}=\frac{\Omega}{2}\sigma_{x}(e^{i\eta(a^\dagger+a)}e^{-i(\omega_Lt+\phi_L)}+{\rm h.c.}),
\end{eqnarray}
where the Rabi frequency $\Omega$ represents the coupling strength between the laser field and the two-level atomic transition and $\eta$ is the Lamb-Dicke parameter. A transformation with $U_0=e^{-iH_0t}$ into the interaction picture yields
\begin{eqnarray}
H_{int}&=&\frac{\Omega}{2}e^{-i(\omega_L-\omega_0)t+\phi_L}\sigma_+\nonumber\\
&&\times[1+i\eta(a^\dagger e^{i\omega_mt}+ae^{-i\omega_mt})]+{\rm h.c.},
\end{eqnarray}
after the rotating wave approximation. When we apply the laser field resonant with the atomic frequency ($\omega_L=\omega_0$), the carrier transition ($(\Omega/2)\sigma_+e^{i\phi_L}+{\rm h.c.}$) is driven. When the laser detuned by negative motional frequency ($\omega_L=\omega_0-\omega_m$) is applied, the red sideband transition is driven ($(\Omega\eta/2)\sigma_+ae^{i\phi_L}+{\rm h.c.}$).

The above analysis is based on a two-level system. By utilizing the multi-level structures of a ${}^{40}{\rm Ca}^+$ ion, we can realize a six-level spin-3/2 system, as shown in Fig.~1(c). We identify the six energy levels with phonon number $n$ as the spin-up and spin-down components of sites $A_n, B_n$, and $C_n$ respectively. The annihilation operators of these sites are denoted as $a_n, b_n$ and $c_n$. We simultaneously apply four lasers to drive the carrier transition and four lasers to drive the red sideband transition between different energy levels. Then we obtain
\begin{eqnarray}
H&=&\frac{J}{2}\sum_{n\geq0}b_n^\dagger U_1a_n+c_n^\dagger U_3a_n+\sqrt{n+1}a^\dagger_{n+1}U_2b_n\nonumber\\
&&+\sqrt{n+1}a^\dagger_{n+1}U_4 c_n+{\rm h.c.},
\end{eqnarray}
which is the experimental Hamiltonian we used. By controlling the magnitude and the phase of the lasers, the gauge field $U_i\in SU(2)$ is synthesized on the link. Note that this Hamiltonian differs from the perfectly translational invariant Hamiltonian in Eq.~\eqref{eq1} by an extra factor $\sqrt{n+1}$ \cite{Leibfried03}. However, the evolution along the upper path and the lower path of a plaquette acquires the same factor. Therefore, apart from the speed up for the evolution, it has no influence for the interference phenomena we are interested in.

\section{Experimental setup, readout sequence and data processing}

To realize the rhombic lattice with $SU(2)$ gauge field, we trap a single ${}^{40}{\rm Ca}^+$ ion in a linear Paul trap with the ambient magnetic field of 0.538 mT. We use the axial motion of the ion as the quantum oscillator. The Lamb-Dicke parameter is $\eta=0.092$. After performing Doppler cooling and the resolved sideband cooling, the axial motion of the ion is cooled close to the ground state. The population of $S_{1/2}$ state of the ion can be read out through the $S_{1/2}-P_{1/2}$ cycling transition. When the valence electron is in the $S_{1/2}$ state, photons are scattered and collected by a photomultiplier tube, whereas if the electron is in the $D_{5/2}$ state, no photons are scattered. A laser at 866 nm repumps the ion via $P_{1/2}-D_{3/2}$ during electron shelving, thus closing the fluorescence excitation cycle \cite{bible98}.

\begin{figure}[t]
\centering
\includegraphics[width=1.\linewidth]{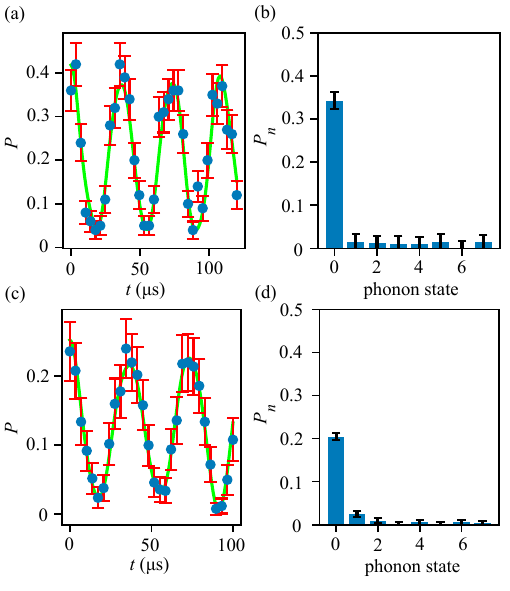}
\caption{Examples of the blue sideband transition curves and the reconstruction of the phonon states. $p$ is the population of the ion bright state. $p(n)$ is the population of the phonon state $|n\rangle$ obtained from the fitting. (a)(b) Each data point is the mean result from 100 repetitions of the experiment. (c)(d) Each data point is the mean result from 400 repetitions of the experiment. We slightly shorten the total time of the sideband transition in (c)(d) to increase the number of data points in one period. Error bars stand for 1 standard deviation. }
\label{figS1}
\end{figure}

We use four states in the $D_{5/2}$ manifold and all two states in the $S_{1/2}$ manifold to encode the lattice sites (Fig.~1(c)). In order to reconstruct the phonon state of all these six energy levels, we conduct the following shelving sequences.
\begin{enumerate}
\item To detect the population of the $|S_{1/2},m=\pm1/2\rangle$ state, we shelve the population of the $|S_{1/2},m=\mp1/2\rangle$ state to the $|D_{5/2},m=\mp5/2\rangle$ state.
\item To detect the population of the $|D_{5/2},m=\pm1/2,\pm3/2\rangle$ state, we first shelve the population of the $|S_{1/2},m=-1/2\rangle$ state to the the $|D_{5/2},m=-5/2\rangle$ state. Then we transfer the population of the state to be measured to the $|S_{1/2},m=+1/2\rangle$ state by driving a $\pi$ pulse.
\end{enumerate}
After the above shelving sequences, we drive the blue sideband transition for varied time $t$ ($|S_{1/2},m=-1/2\rangle-|D_{5/2},m=-5/2\rangle$ is used for the read out of $|S_{1/2},m=-1/2\rangle$ state, otherwise $|S_{1/2},m=+1/2\rangle-|D_{5/2},m=+5/2\rangle$ is used). Then we drive the 397 nm cycling transition and obtain the bright state population of the ion.

For a state $\sum_{n=0}\sqrt{p(n)}|\downarrow,n\rangle$, where $|\downarrow\rangle$ denote the bright state after shelving, the evolution for the blue sideband transition follows the form
\begin{equation}
P(\downarrow,t)=\frac{1}{2}\sum_{n=0}p(n)(1+\cos(\Omega\frac{e^{-\eta^2/2}}{\sqrt{n+1}}\eta L_n^1(\eta^2) t),
\end{equation}
where $P(\downarrow,t)$ is the population of the bright state of the ion at time $t$, $L_n^\alpha(X)$ is the generalized Laguerre polynomial, and $\Omega$ is the Rabi frequency. To fit the phonon population, we first drive the blue sideband transition immediately after the state preparation, which gives us the sideband Rabi frequency $\Omega\eta$. We fix this parameter during the fitting. Then, a fitting to the blue sideband transition curves allows us to reconstruct the population of the phonon state $p(n)$ (Fig.~\ref{figS1}). The phonon state of every spin level can be obtained through the above shelving sequences. Thus we obtain the population of all the lattice sites.

\begin{figure}[t]
\centering
\includegraphics[width=1.\linewidth]{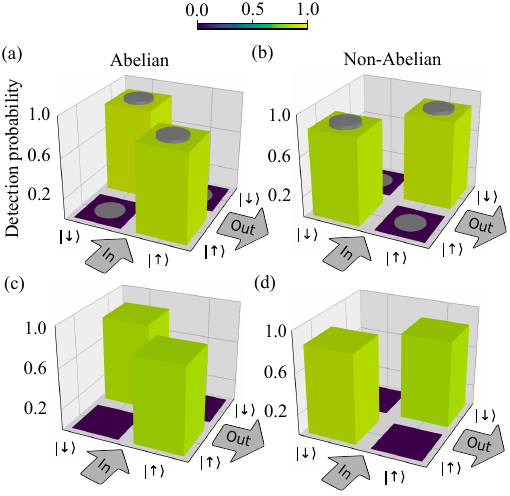}
\caption{Tomography of the $n=0$ plaquette. The Wilson loop is obtained from the absolute value of the trace of the matrix. The $\ket\uparrow (\ket\downarrow)$ state is the spin-up (spin-down) component of site $A_0$. (a)(b) Experimental results. (c)(d) Numerical simulations. The left and the right panel correspond to the Abelian gauge field and the non-Abelian gauge field set in Fig.~2, respectively. The gray cylinders correspond to 1 standard deviation.}
\label{figS2}
\end{figure}

To fit the blue sideband curves, we need to choose an appropriate phonon fitting cutoff $n_{max}$. A too low $n_{max}$ results in the distortion of the fitting result, while a too high $n_{max}$ will make the fitting error irrational large. In our experiment, we choose $n_{max}=7$, which is large enough to reconstruct the phonon state during the evolution \cite{Wang24}.

To measure the dynamical evolution diagram, each data point in the blue sideband curves is the mean measured result based on 100 repetitions of the experiment. To quantitatively analyze the localization (Fig.~2(f) and Fig.~4(c)), we increase the repetition to 400 times per data point to acquire higher precision.

When calibrating the frequency of each laser, we simultaneously turn on all eight lasers, while set the other seven lasers detuning. Thus the AC stark shift is mostly compensated during the calibration \cite{Meth25}. 

The experimental imperfections will cause the measured results to slightly deviate from the ideal results. One of the main imperfections is the magnetic jitter of the environment during the experimental sequence, which makes the lasers detuned from the transition, perturbs the caging condition, and leads to minor population leakage with quantity $\sim 0.1$, e.g. in Fig.~1(a-c). Another error is the imperfect ground state cooling, which slightly reduces the initial phonon number in the $n=0$ sites, e.g., causing the population $P_0$ in Fig.~2(f) under the caging condition less than the ideal value 1.

\begin{figure}[t]
\centering
\includegraphics[width=1.\linewidth]{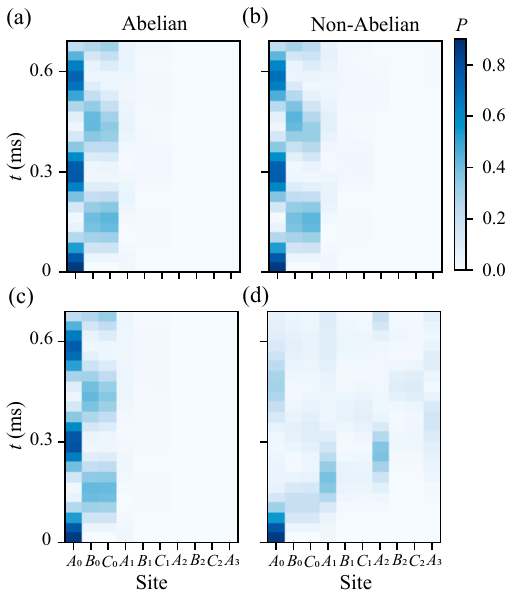}
\caption{Numerical simulated dynamical evolution diagrams under the Abelian (left panel) and the non-Abelian gauge field (right panel). The initial state is set as an out-of-phase superposition state (a)(b) and an in-phase superposition state (c)(d) at site $A_0$, respectively.}
\label{figS3}
\end{figure}

\section{Measurement of the Wilson loop}

We use the Wilson loop $|{\rm Tr}[U(\gamma)]|$ to justify whether the synthetic gauge field is Abelian or non-Abelian \cite{Neef23, Goldman14}. In the main text, we directly calculate the Wilson loop through the coupling. We also experimentally measure the Wilson loop for the coupling set in Fig.~2. We initialize the system at one of the two internal spins of site $A_0$, successively turn on the couplings of the four links in the plaquette for a duration of $\pi$ time, and finally measure the population of the two internal spins. The measured Wilson loop is 1.728(0.07) for the Abelian gauge field and 0.010(0.013) for the non-Abelian gauge field (Fig.~\ref{figS2}). For the Abelian gauge field, the discrepancy between the measured value and the ideal value 2 is mainly due to the magnetic jitter during the experiment and the noise of the laser strength.

\section{Numerical simulation}

The numerical simulations are performed by solving the Lindblad master equation to account for the influence of the environment
\begin{eqnarray}
\dot{\rho}=-\frac{i}{\hbar}[H,\rho]+\frac{1}{2}\sum_n[2C_n\rho C_n^\dagger-\rho C_n^\dagger C_n-C_n^\dagger C_n\rho]\nonumber\\
\end{eqnarray}
where $C_n$ is the collapse operator \cite{Matsos24}. The collapse operators we consider include the phonon decoherence terms ($\sqrt{\Gamma_1}a^\dagger$ and $\sqrt{\Gamma_1}a$) and the spin decoherence term ($\sqrt{\Gamma_2}\sigma_z$, here $\sigma_z$ consists of the $D_{5/2}$ state population operators minus the $S_{1/2}$ state population operators). The heating rate $\Gamma_1$ is 100 Hz and the spin decoherence rate $\Gamma_2$ is 200 Hz. The magnetic jitter is considered by adding a constant frequency detuning, which equals about 220 Hz, to the Hamiltonian. The imperfect ground state cooling is considered by setting the initial phonon state in the simulations according to the measured initial state. The simulation cutoff of the phonon number is set at $N=8$.

The simulation results of the AB caging under the Abelian gauge field and the non-Abelian gauge field, corresponding to Fig.~2(a-d) in the main text, are shown in Fig.~\ref{figS3}.

\section{Analysis of the caging condition}

In the main text, we experimentally explore the properties of the non-Abelian AB caging, including the initial-state dependence, the second-order caging and the asymmetry caging, and we claim that they are unique for the non-Abelian AB caging. Here we theoretically prove that these phenomena are unique to the non-Abelian gauge field.

If not, suppose the synthetic gauge field is Abelian, that is, the Wilson loop equals 2. Then $U_3^\dagger U_4^\dagger U_2U_1\propto \hat{\bm{1}}$. Thus $U_2U_1=e^{i\theta}U_4U_3$. The interference matrix becomes
\begin{eqnarray}
I=&\frac{1}{2}(U_4U_3+U_2U_1)\nonumber=\frac{1+e^{i\theta}}{2}U_4U_3.
\end{eqnarray}
Therefore, there are only two possibilities. When $\theta=\pi$, we have $I=0$, which results in a trivial symmetric first-order caging independent of the initial state, known as the Abelian AB caging. Otherwise, if $\theta\neq\pi$, we have
\begin{equation}
I^m=\left(\frac{1+e^{i\theta}}{2}\right)^m(U_4U_3)^m.
\end{equation}
Its coefficient is nonzero. Moreover the $(U_4U_3)^m|\psi\rangle$ is nonzero for any $|\psi\rangle$ (Note that the unitary operator is norm-preserving). Therefore $I^m|\psi\rangle$ is always nonzero, so does the Hermitian conjugate $(I^{\dagger})^{m}|\psi\rangle$. Therefore, no caging phenomena can occur, which completes our proof.

\end{document}